\documentclass[seceq]{ptptex}

\usepackage{graphicx}
\usepackage{epsfig}

\newcommand{\fpi}{f_\pi}
\newcommand{\mpi}{m_\pi}
\newcommand{\gev}{\,{\rm GeV}}

\newcommand{\fm}{\,{\rm fm}}
\newcommand{\ra}{\rightarrow}

\newcommand{\sigNNpi}{\sigma_{NN}^\pi}
\newcommand{\sigNDpi}{\sigma_{N\Delta}^\pi}
\newcommand{\sigTAD}{\sigma_{\rm tad}^\pi}

\title{
Towards a Connection Between Nuclear Structure and QCD%
}

\author{
Anthony W.\ \textsc{Thomas}$^{1,2}$, Pierre A.\ M.\ \textsc{Guichon}$^3$,
Derek B.\ \textsc{Leinweber}$^2$ \\ and Ross D.\ \textsc{Young}$^2$ 
}

\inst{
$^1$Jefferson Lab, 12000 Jefferson Ave., Newport News VA 23606 USA \\
$^2$Centre for the Subatomic Structure of Matter and \\
Department of Physics, University of Adelaide, Adelaide SA 5005, Australia \\
$^3$SPhN-DAPNIA, CEA Saclay, F91191 Gif sur Yvette, France
}

\abst{
As we search for an ever deeper understanding of the structure of 
hadronic matter one of the most fundamental questions is whether or 
not one can make a connection to the underlying theory of the strong 
interaction, QCD. We build on recent advances in the chiral extrapolation 
problem linking lattice QCD at relatively large ``light quark'' masses to 
the physical world to estimate the scalar polarizability of the nucleon. 
The latter plays a key role in modern relativistic mean-field descriptions of 
nuclei and nuclear matter (such as QMC) 
and, in particular, leads to a very natural 
saturation mechanism. We demonstrate that the value of the scalar 
polarizability extracted from the lattice data is consistent with that needed 
for a successful description of nuclei within the framework of QMC. In 
a very real sense this is the first hint of a direct connection between 
QCD and the properties of finite nuclei. 
}

\begin{document}

\maketitle

\section{Introduction}
Over the century since Rutherford made the first discovery concerning 
nuclear structure, understanding the mechanism for nuclear saturation 
has been a central concern. Within the traditional framework of many-body 
theory based on two-body potentials, strong short-distance repulsion was 
the key. The importance of relativity was made clear within 
quantum-hadrodynamics (QHD), with the scalar density of nucleons (which 
determines the attractive scalar, $\sigma$-meson, coupling) growing more 
slowly than the density (which in turn determines the repulsive  
vector, $\omega$-meson, coupling). 

Since the discovery of QCD as the fundamental theory of the strong interaction, 
numerous attempts have been made to derive the NN force within  
quark models. Fewer attempts have been made to understand nuclear structure 
at the quark level~\cite{Benesh:2003fk,Smith:ak}. 
The quark-meson coupling (QMC) 
model~\cite{Guichon:1987jp,Saito:ki,Saito:1994kg,Guichon:1995ue} 
stands between the traditional meson-exchange picture and the 
hard core quark models. It is a mean field model in the sense of QHD but 
with the $\sigma$ and $\omega$ mesons coupling to confined quarks, 
rather than to point-like nucleons. In the simple case of infinite 
nuclear matter, after a considerable amount of work, one finds that the 
only effect of the internal, quark structure of the nucleons is that 
the $\sigma$-nucleon coupling becomes density dependent. That is, the 
nucleon effective mass takes the form:
\begin{equation}
M_N^* = M_N - g_{\sigma} (\sigma) \sigma
\label{eq:1}
\end{equation}
where $g_{\sigma}(\sigma) \approx g_{\sigma} + g_{\sigma \sigma} \sigma$, 
$g_{\sigma}$ is the coupling at zero density and
$g_{\sigma \sigma}$ is known as the scalar 
polarizability~\cite{Guichon:1987jp,Guichon:1995ue,Bentz:2001vc,Chanfray:2003rs}. 

If the MIT bag model is used to 
describe the quark confinement (as in the original QMC model) 
the scalar polarizability is negative, that 
is it tends to oppose the applied scalar field. Indeed, one finds
$g_{\sigma \sigma} = - 0.11 g_\sigma R$, with $R$ the bag radius.
A similar result is found in the NJL model\cite{Bentz:2001vc}, 
once it is modified to 
simulate the effect of confinement~\cite{Hellstern:1997nv} 
-- a change which also 
resolves the long standing question of chiral collapse in the NJL model.

The crucial point is now that a negative scalar polarizability (associated 
with a readjustment of the internal, quark structure of the nucleon 
in response to an applied scalar field) is all that is needed to lead to 
the saturation of the binding energy of nuclear matter as a function of 
density. In contrast with the Walecka model (QHD)~\cite{Serot:1997xg}, 
where the $\sigma$-nucleon 
coupling really is a constant, the mean scalar and vector fields at 
nuclear matter density are much smaller because one is not relying on 
a difference between $\bar{\psi}_N \psi_N$ and $\psi_N^\dagger \psi_N$ 
arising from relativistic motion of the bound nucleons.
 
While the situation just described is somewhat simplistic as a 
description of nuclear saturation, it is certainly sufficient. Moreover, 
it has recently been shown that a systematic expansion of the QMC model 
as a function of density allows a direct comparison with the widely used
Skyrme effective forces~\cite{Guichon:2004xg}. 
In that framework one can also include 
anti-symmetrization at the nucleon level, leading to new values of the $\sigma$ 
and $\omega$ coupling constants. With these values the expansion of the effective model
as an effective nuclear force yields agreement at the 10\% level. This provides 
a remarkable indication of the role of confined quarks in even low energy nuclear 
physics.

\section{The Status of Chiral Extrapolation for the Mass of the Nucleon}

The challenge of solving a strongly coupled
{}field theory with spontaneous symmetry
breaking is central to understanding the strong interaction within the 
Standard Model. 
It will be quite a few years before full QCD simulations can be
performed at physical light quark masses -- with computation time
scaling like $\sim 1/m_q^{3.6}$ \cite{Lippert:zq}, 
quark masses are typically restricted to
above $50\,{\rm MeV}$ for accurate calculations with improved quark
and gluon actions. The exception to this is to use staggered fermions, 
which have shown some results at $25\,{\rm MeV}$ \cite{Bernard:2001av}, 
but this approach has a
number of technical problems, including 
multiple pions (extra tastes) -- which  
complicate the chiral extrapolation problem which we explore next.

Given that one cannot directly calculate hadron properties at realistic 
quark masses, if one wishes to make any connection with experimental 
data it is necessary to find an appropriate method to extrapolate the 
properties calculated at large {\em light-quark} mass to the physical value.
This is the chiral extrapolation problem, which is complicated by
spontaneous chiral symmetry breaking in QCD. As we shall explain, 
this problem has recently been solved by a careful reformulation of
the effective field theory.

The power of effective field theory for strongly interacting systems
has been apparent for more than 50 years, beginning with Foldy's
inclusion of the anomalous magnetic moment of the proton in a
derivative expansion of the electromagnetic coupling of photons to
baryons. Of course, the fact that nucleon form factors can be
approximated by a dipole with mass parameter $M^2=0.71\,{\rm GeV}^2$
means that the radius of convergence of the formal expansion is less
than $0.84\,{\rm GeV}$. In the last twenty years our understanding of the
symmetries of QCD have led to a systematic expansion of hadron
properties and scattering amplitudes about the limits $\mpi=0$ and
$q=0$, known as chiral perturbation theory. For the pseudoscalar
mesons, which are rigorously massless in the limit where the $u$, $d$
and $s$ quark masses are zero (the chiral $SU(3)$ limit), this
approach has been especially successful. Although, even
here, serious questions have been raised over whether the formal
expansion is really convergent for masses as large as $m_s$
\cite{Li:1971vr,Hatsuda:tt,Durr:2002zx}.

{}For our immediate interest in the scalar polarizability of the nucleon,
we shall concentrate on understanding the dependence of $M_N$ on light quark
mass $m_q$. Given the Gell-Mann--Oakes--Renner (GMOR) relation,
$\mpi^2\propto m_q$, we prefer to use $\mpi^2$, rather than $m_q$, as the
measure of explicit chiral symmetry breaking in the expressions
below. (We stress, however, that wherever the GMOR relation
{}fails, e.g.~for $\mpi^2>0.8\,{\rm GeV}^2$ or for $\mpi\to 0$ in QQCD
(where there are logarithmic chiral corrections to GMOR), one should revert to
an expansion written explicitly in terms of $m_q$.) The formal
chiral expansion for $M_N$, in terms of $\mpi$, about the $SU(2)$
chiral limit in full QCD is:
\begin{eqnarray}
M_N &=& a_0 + a_2 \mpi^2 + a_4 \mpi^4 + a_6 \mpi^6 + \ldots \nonumber\\
    & &  + \sigma_{{\rm N} \pi}(\mpi) 
         + \sigma_{\Delta \pi}(\mpi) \, ,
\label{eq:eftexp}
\end{eqnarray}
where $\sigma_{{\rm B} \pi}$ is the self-energy arising from a $B\pi$
loop (with $B$ = $N$ or $\Delta$). These $N$ and $\Delta$ loops
generate the leading and next-to-leading non-analytic (LNA and NLNA)
behaviour, respectively. Note that for the present we have omitted the 
tadpole term which also contributes at NLNA order. It adds 
complications which are unimportant for the present discussion -- for 
details we refer to Sec. 2.3 and to Ref.\cite{Leinweber:2003dg}.

These loop integrals contain ultraviolet divergences which
require some regularisation prescription. The standard approach is to
use dimensional regularisation to evaluate the self-energy integrals.
Under such a scheme the $NN\pi$ contribution simply becomes
$\sigma_{{\rm N} \pi}(\mpi)\to c_{\rm LNA}\mpi^3$ and the
analytic terms, $a_0$ and $a_2\mpi^2$, undergo an infinite
renormalisation. The $\Delta$ contribution produces a logarithm, so
that the complete series expansion of the nucleon mass about $\mpi= 0$
is:
\begin{equation}
M_N = c_0 + c_2 m_\pi^2 + c_4 m_\pi^4 + c_{\rm LNA} m_\pi^3 
        + c_{\rm NLNA} m_\pi^4 \log m_\pi + \ldots\, ,
\label{eq:drexp}
\end{equation}
where the $a_i$ have been replaced by the renormalised (and
finite) parameters $c_i$.

The coefficients of the low-order, non-analytic
contributions are known~\cite{Pagels:se,Gasser:1988rb,Lebed:1994yu}:
\begin{equation}
c_{\rm LNA}=-\frac{3}{32\,\pi\,\fpi^2}\,g_A^2\, , \hspace{5mm} 
c_{\rm NLNA}=\frac{3}{32\,\pi\,\fpi^2}\,\frac{32}{25}g_A^2\,
\frac{3}{4\,\pi\,\Delta}\, .
\end{equation}
Although strictly one should use values in the chiral limit, we take
the experimental numbers with $g_A=1.26$, $\fpi=0.093\gev$, the
$N-\Delta$ mass splitting, $\Delta=0.292\gev$
and the mass scale associated with the logarithms will be taken to be
$1\gev$.

We stress that whereas Eq. (\ref{eq:drexp}) was derived in the limit
$\mpi/\Delta \ll 1$, at just twice the physical pion mass this ratio
approaches unity. Mathematically the region $m_\pi \approx \Delta$ is
dominated by a square root branch cut which starts at $m_\pi =
\Delta$. Using dimensional regularisation this takes the form
\cite{Banerjee:1996wz}: 
{\scriptsize
\begin{equation}
\frac{-3}{32\,\pi\,\fpi^2}\,\frac{32}{25}g_A^2
\frac{1}{2\pi}\left[ (2\Delta^3 - 3\mpi^2\Delta )
  \log\left(\frac{\mpi^2}{\mu^2}\right)
- 2 (\Delta^2-\mpi^2)^{\frac{3}{2}}
\log\left(\frac{\Delta-\sqrt{\Delta^2-\mpi^2}}{\Delta+\sqrt{\Delta^2-\mpi^2}}\right)
\right]
\label{eq:log}
\end{equation}
}
{}for $m_\pi < \Delta$, while for $m_\pi > \Delta$ the second logarithm
becomes an arctangent. Clearly, to access the higher quark masses in the
chiral expansion, currently of most relevance to lattice QCD,
one requires a more sophisticated expression than
that given by Eq.~(\ref{eq:drexp}).

\subsection{Issues of convergence}
{}Forgetting for a moment the issues surrounding the $\Delta \pi$ cut, 
we note that the formal
expansion of the $N\, \ra\, N\,\pi\, \ra\, N$ self-energy integral,
$\sigma_{{\rm N} \pi}$, has been shown to have poor convergence
properties. Using a sharp, ultra-violet cut-off, Wright showed
\cite{SVW} that the series expansion, truncated at ${\cal O}(\mpi^4)$,
diverged for $m_\pi > 0.4$ GeV. This already indicated that the series
expansion motivated by dimensional regularisation would have a slow
rate of convergence. A similar conclusion had been reached somewhat
earlier by Stuckey and Birse \cite{Stuckey:1996qr}.
\begin{figure}[t]
\begin{center}
\includegraphics[width=.8\textwidth]{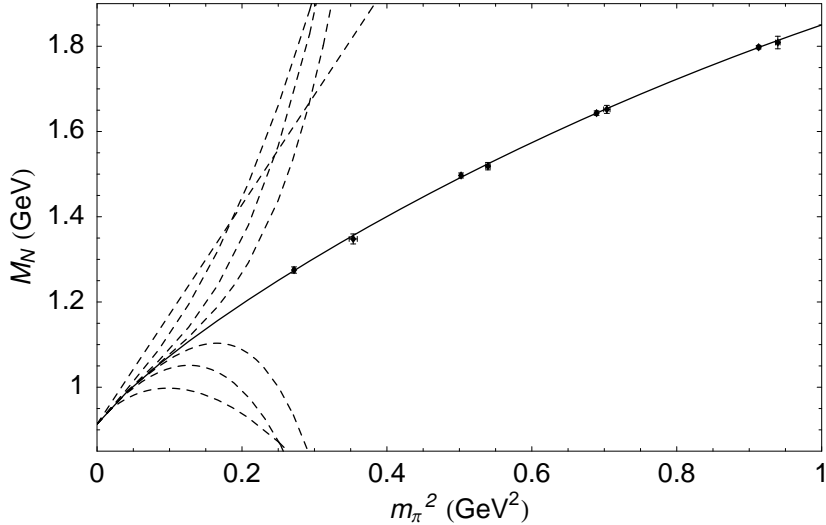}
\end{center}
\caption[]{Fits to lattice data with dipole regulator for the
self-energy loop $N\to N\pi$ (neglecting the $N\to \Delta \pi$ term for
simplicity). The dashed lines show the expansion up to successive orders
in $m_\pi$, from $m_\pi^2$ up to $m_\pi^8$ -- these alternate about the
solid curve as we go through even and odd terms.}
\label{fig:expand}
\end{figure}

In considering the convergence of the truncated series,
Eq.~(\ref{eq:drexp}), it is helpful to return to 
the general form from which it was derived, namely Eq.~(\ref{eq:eftexp}). The
dimensionally regulated approach requires that the pion mass should
remain much lighter than every other mass scale involved in the
problem. This requires that $\mpi/\Lambda_{\chi {\rm SB}}\ll 1$ (and
$\mpi/\Delta\ll 1$ if we use the simple logarithm in 
Eq.(\ref{eq:drexp}) rather than the full cut structure of 
Eq.(\ref{eq:log}) ).

An additional mass scale, which we address in detail below, 
is set by the physical
extent of the source of the pion field \cite{Thomas:1982kv}. 
This scale, which is of order
$R_{SOURCE}^{-1}$, corresponds to the transition between the rapid,
non-linear variation required by chiral symmetry and the smooth, {\em 
constituent-quark} like mass behaviour observed in lattice
simulations at larger quark mass. An alternative to dimensional
regularisation is to regulate Eq.~(\ref{eq:eftexp}) with a
finite ultra-violet cut-off in momentum space.

General considerations in effective field theory, as discussed for example by
LePage~\cite{Lepage:1997cs}, suggest that one should 
{\em not} take the cut-off to $\infty$. That is, one should not use
dimensional regularization. Nor should one attempt to determine the
``true'' cut-off of the theory. Rather, one should choose a cut-off scale
somewhat below the place where the effective theory omits 
essential physics and use data to constrain the renormalization
constants of the theory and hence eliminate the
dependence on that cut-off as far as
possible.

The issue is then what mass scale determines the upper limit beyond
which the effective field theory is applicable. This is commonly taken
to be $\Lambda_{\chi PT} \sim 4 \pi f_\pi \sim 1$ GeV. Unfortunately
this is incorrect for baryons. In the context of nuclear physics it has
long been appreciated that nuclear sizes could never be derived from
naive dispersion relation considerations of nearest t-channel poles --
anomalous thresholds associated with the internal structure of nuclei
dominate. Similarly, the size of a baryon is determined by
nonperturbative QCD beyond the scope of chiral perturbation theory. The
natural scale associated with the size of a nucleon is the inverse of
its radius or a mass scale $R_{SOURCE}^{-1} \sim $ 0.2 to 0.5 GeV -- far below
$\Lambda_{\chi PT}$. In the context of effective field theory it 
is inconsistent to keep loop
contributions from momenta above this scale! 

As a result of these considerations, we are led to regulate the
radiative corrections which give rise to the leading and next-to-leading
non-analytic contributions to the mass of the nucleon using a finite
range regulator with a mass in the range $R_{SOURCE}^{-1}$. As far as
possible residual dependence on the specific choice of mass scale (and
form of regulator function) will be eliminated by fitting the
renormalization constants to nonperturbative QCD -- in this case data
obtained from lattice QCD simulations. The quantitative success of
applying the method is to be judged by the extent to which physical
results extracted from the analysis are indeed independent of the
regulator (over a physically sensible range). We refer to this approach 
as {\em finite range regularization}, FRR.

\subsection{Analysis of lattice QCD data}
In our analysis we have taken as input both the physical 
nucleon mass and recent lattice QCD
results of the CP-PACS Collaboration \cite{AliKhan:2001tx} and the
JLQCD Collaboration \cite{Aoki:2002uc}. This enables us to constrain
an expression for $M_N$ as a function of the quark mass.
The lattice results of Ref.~\cite{AliKhan:2001tx} have been obtained
using improved gluon and quark actions on fine, large volume lattices
with high statistics.
These simulations were performed using an
Iwasaki gluon action \cite{Iwasaki:1985we} and the mean-field improved
clover fermion action. In this work we concentrate on only those
results with $m_{\rm sea} = m_{\rm val}$ and the two largest values of
$\beta$ (i.e., the finest lattice spacings $a\sim 0.09$ -- $0.13\fm$).
We use just the largest volume results of Ref.~\cite{Aoki:2002uc}, where
simulations were performed with non-perturbatively improved Wilson
quark action and plaquette gauge action. The lattice volumes and
spacings are similar for the two data sets.

We chose to set the physical scale at each quark mass 
using the UKQCD method \cite{Allton:1998gi}. That is, we use the
Sommer scale $r_0=0.5\fm$ \cite{Sommer:1994ce,Edwards:1998xf}. This
choice is ideal in the present context because the static quark
potential is insensitive to chiral physics. This ensures that the
results obtained represent accurate estimates of the continuum,
infinite-volume theory at the simulated quark masses. The lattice data
lies in the intermediate mass region, with $\mpi^2$ between $0.3$ and
$1.0\gev^2$.

Based on Eq. (\ref{eq:eftexp}) we evaluate the loop integrals
$\sigma_{{\rm N} \pi} \equiv c_{LNA} I_\pi(m_\pi,\Lambda)$ and 
$\sigma_{\Delta \pi} \equiv c_{NLNA} 
I_{\pi\Delta}(m_\pi,\Lambda)$, where: 
\begin{equation}
I_\pi = \frac{2}{\pi} \int_0^\infty dk\, \frac{k^4\, u^2(k)}{k^2+m^2}\,
\end{equation}
and 
\begin{equation}
I_{\pi\Delta} = -\frac{8 \Delta}{3} \int_0^\infty dk\,
\frac{k^4\, u^2(k)}{\sqrt{k^2+m^2}(\Delta+\sqrt{k^2+m^2})}\, .
\end{equation}
We use four different functional
forms for the finite-ranged, ultra-violet vertex regulator, $u(k)$ --- 
namely the sharp-cutoff (SC), $\theta(\Lambda-k)$; monopole (MON),
$\Lambda^2/(\Lambda^2+k^2)$; dipole (DIP),
$\Lambda^4/(\Lambda^2+k^2)^2$; and Gaussian (GAU),
$\exp(-k^2/\Lambda^2)$. Closed expressions for these integrals in the
first three cases are given in the Appendix of Ref.~\cite{Young:2002ib}.
Provided one regulates the effective field theory below the point
where new short distance physics becomes important, the essential
results will not depend on the cut-off scale \cite{Lepage:1997cs}. We
use knowledge learned in Ref.~\cite{Young:2002ib} as a guide for the
appropriate scales for each of the regulator forms. 

\begin{figure}[htb]
\begin{center}
\includegraphics[width=.8\textwidth]{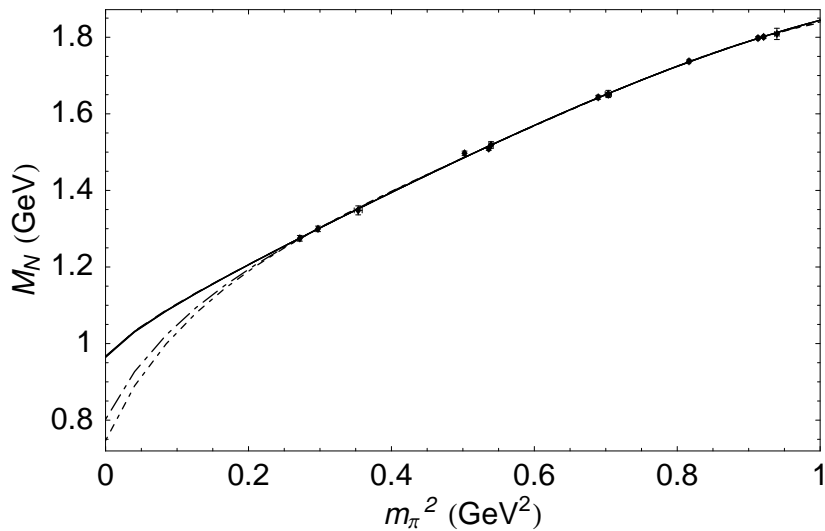}
\caption{Extrapolation of lattice data for various regularisations
prescriptions, without the tadpole term. 
The four {\bf indistinguishable,} solid curves show
the fits based on four different FRRs. The short dash-dot curve
corresponds to the dimensionally regulated case and the long dash-dot 
curve to that where the branch
point in Eq.~(\ref{eq:log}) is included in dimensional regularisation.
\label{fig:extrap}}
\end{center}
\end{figure}

We have fit the set of lattice data discussed above with the chiral expansions
based on six different regularisation prescriptions. In all schemes we
allowed the coefficients of the analytic terms up to $\mpi^6$ to be
determined by the lattice data. The regulator masses are once again
fixed to their preferred values. The best fits, shown in
Fig.~\ref{fig:extrap},
serve to highlight the remarkable agreement between all finite-range
parameterisations. This demonstrates that the extrapolation can be
reliably performed using FRR with negligible dependence on the choice
of regulator.
\begin{figure}[htb]
\begin{center}
\includegraphics[width=.8\textwidth]{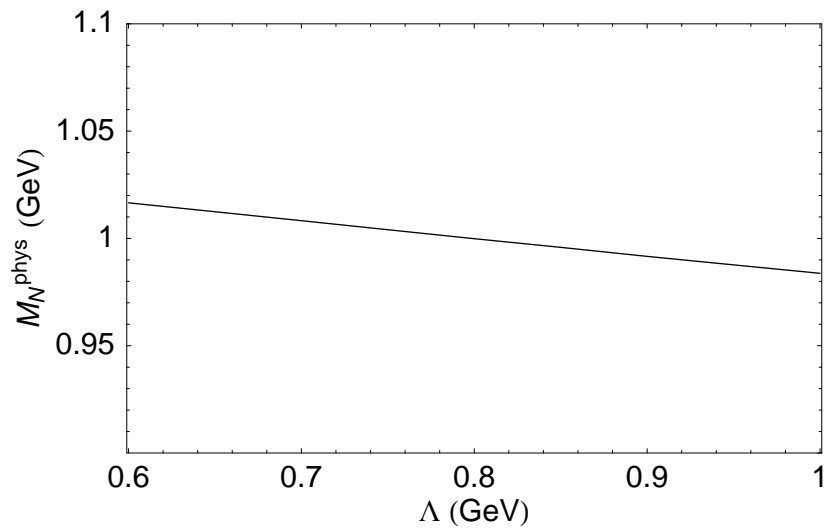}
\caption{Illustration of the relatively weak dependence of the extrapolated nucleon 
mass on the choice of dipole regulator mass,
$\Lambda$.
\label{fig:lamdep}}
\end{center}
\end{figure}

In view of the general discussion presented earlier, it is important to check
the residual dependence on the choice of cutoff scale
$\Lambda$. The resulting variation of the
extrapolated nucleon mass, at the physical pion mass, for dipole
masses ranging from $0.6$ to $1.0\gev$ is shown in
Fig.~\ref{fig:lamdep}. We see that the
residual uncertainty introduced by the cutoff scale is less than 2\%, which 
is insignificant compared to the statistical error in
extrapolating such a large distance. With the present data this statistical 
error is found to be around 13\%.

\subsection{Complete analysis of the nucleon mass}
Here we summarise the results of the most recent and complete analysis of 
Ref.~\cite{Leinweber:2003dg}. Including the tadpole term, the nucleon mass
has the formal expansion:
\begin{equation}
M_N = a_0 + a_2 m_\pi^2 + a_4 m_\pi^4 + a_6 m_\pi^6 
     + \sigNNpi + \sigNDpi + \sigTAD.
\label{eq:formal}
\end{equation}
As explained earlier, within dimensional regularisation this leads to 
a truncated power series for the chiral expansion,
\begin{equation}
M_N = c_0 + c_2 m_\pi^2 + c_{\rm LNA} m_\pi^3 + c_4 m_\pi^4 
     + c_{\rm NLNA} m_\pi^4 \ln \frac{m_\pi}{\mu}
	+ c_6 m_\pi^6 + \ldots \, ,
\label{eq:drexp2}
\end{equation}
where the bare parameters, $a_i$, have been replaced by the finite,
renormalised coefficients, $c_i$. Through the chiral logarithm one has
an additional mass scale, $\mu$, but the dependence on this is
eliminated by matching $c_4$ to ``data'' (in this case lattice QCD).
We work to fourth order in the chiral expansion and include the next
analytic term to compensate short distance physics contained in the
NLNA loop integrals as suggested in Ref.~\cite{Donoghue:1998bs}.

On the other hand, the systematic FRR expansion
of the nucleon mass is:
\begin{eqnarray}
M_N &=& a_0^\Lambda + a_2^\Lambda m_\pi^2 + a_4^\Lambda m_\pi^4 
        + a_6^\Lambda m_\pi^6 \nonumber 
+ \sigNNpi(m_\pi, \Lambda) \\
    & & 
        + \sigNDpi(m_\pi, \Lambda) + \sigTAD(m_\pi, \Lambda) ,
\label{eq:finite}
\end{eqnarray}
where the dependence on the {\it shape} of the regulator is implicit.
The dependence on the value of $\Lambda$ and the choice of regulator
is eliminated, to the order of the series expansion, by fitting the
coefficients, $a_n^\Lambda$, to lattice QCD data. The clear indication
of success in eliminating model dependence and hence having found a
suitable regularisation method, is that the higher order coefficients
($a_i^{\Lambda} \, , i \geq 4$) should be small and that the
renormalised coefficients, $c_i$, and the result of the extrapolation
should be insensitive to the choice of ultraviolet regulator.
The forms of the $NN\pi$ and $N \Delta \pi$ contributions were given earlier, 
while the tadpole term has the form:
\begin{equation}
\sigTAD = -\frac{3}{16\pi^2\fpi^2}\, c_2\, \mpi^2
\left\{ \int_0^\infty dk \left(\frac{2 k^2 u^2(k)}{\sqrt{k^2+\mpi^2}}\right) -
  t_0\right\}\, ,
\label{eq:tad}
\end{equation}
where in Eq.~(\ref{eq:tad}) $t_0$, defined such that the term in braces
vanishes at $\mpi=0$, is a local counter term introduced to
ensure a linear relation for the renormalisation of $c_2$.

\begin{figure}[t]
\begin{center}
\epsfig{file=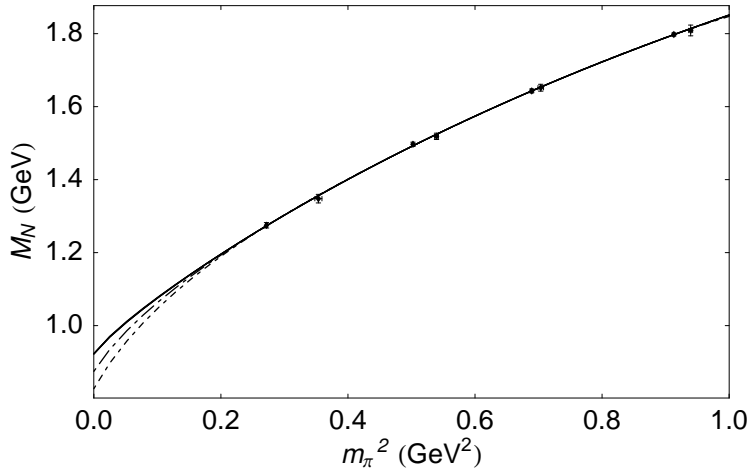,angle=0,width=10cm}
\caption{Fits to lattice data for various ultra-violet 
  regulators including the tadpole term 
  -- from Ref.~\protect\cite{Leinweber:2003dg}.
  The sharp cut-off, monopole, dipole and Gaussian cases are depicted
  by solid lines, {\bf indistinguishable} on this plot. The
  dimensional regularised forms are illustrated by the dash-dot
  curves, with the correct branch point corresponding to the higher
  curve. Lattice data is from Ref.~\protect\cite{AliKhan:2001tx}.}
\label{fig:latfit}
\end{center}
\end{figure}

The result of the fit to the latest two-flavor lattice QCD data for the 
mass of the nucleon is shown in Fig.~\ref{fig:latfit}, with the corresponding
parameters given in Table~\ref{tab:afit}.
It is remarkable that all
four curves based on FRR are indistinguishable on this plot.
Furthermore, we see from Table~\ref{tab:afit} that the coefficient of
$m_\pi^4$ in all of those cases is quite small -- an order of
magnitude smaller than the dimensionally regularised forms.
Similarly, the FRR coefficients of $\mpi^6$ are again much smaller
than their DR counterparts.  This indicates that the residual series,
involving $a_i$, is converging when the chiral loops are evaluated
with a FRR.
\begin{table}
\begin{center}
\begin{tabular}{lcccccc}
Regulator	& $a_0$    & $a_2$    & $a_4$    & $a_6$    & $\Lambda$ & 
$\chi^2/{\rm dof}$\\
\hline
Dim.~Reg.	& $0.827$  & $3.58$   & $3.63$   & $-0.711$ & --     & $0.43$ \\
Dim.~Reg.~(BP)	& $0.792$  & $4.15$   & $8.92$   & $0.384$  & --     & $0.41$ \\
Sharp Cutoff	& $1.06$   & $1.47$   & $-0.554$ & $0.116$  & $0.4$  & $0.40$ \\
Monopole	& $1.74$   & $1.64$   & $-0.485$ & $0.085$  & $0.5$  & $0.40$ \\
Dipole		& $1.30$   & $1.54$   & $-0.492$ & $0.089$  & $0.8$  & $0.40$ \\
Gaussian	& $1.17$   & $1.48$   & $-0.504$ & $0.095$  & $0.6$  & $0.40$
\end{tabular}
\caption{Bare, unrenormalised, parameters extracted from the fits to lattice
data displayed in Fig.~\ref{fig:latfit} -- from 
Ref.~\protect\cite{Leinweber:2003dg}.  All quantities are in
units of appropriate powers of GeV and $\mu=1$ GeV 
in Eq.~(\protect\ref{eq:drexp2}).
}
\label{tab:afit}
\end{center}
\end{table}

The corresponding, physically meaningful, renormalized coefficients are shown in
Table~\ref{tab:cren}, in comparison with the corresponding DR
coefficients found using Eq.~(\ref{eq:drexp2}).  Details of this
renormalisation procedure are given in Ref.~\cite{Young:2002ib}.
\begin{table}
\begin{center}
\begin{tabular}{lccc}
Regulator       & $c_0$        & $c_2$       & $c_4$  \\
\hline
Dim.~Reg.	& $0.827(120)$ & $3.58(50)$  & $3.6(15)$  \\
Dim.~Reg.~(BP)	& $0.875(120)$ & $3.14(50)$  & $7.2(15)$  \\
Sharp cutoff	& $0.923(130)$ & $2.61(66)$  & $15.3(16)$ \\
Monopole	& $0.923(130)$ & $2.45(67)$  & $20.5(30)$ \\
Dipole		& $0.922(130)$ & $2.49(67)$  & $18.9(29)$ \\
Gaussian	& $0.923(130)$ & $2.48(67)$  & $18.3(29)$
\end{tabular}
\caption{Renormalised expansion coefficients in the chiral limit
obtained from various regulator fits to lattice 
data -- from Ref.~\protect\cite{Leinweber:2003dg}. (All quantities
are in units of appropriate powers of GeV.)
Errors are statistical in origin arising from lattice data. Deviations
in the central values indicate systematic errors associated with the
chiral extrapolation.
\label{tab:cren}}
\end{center}
\end{table}
The degree of consistency between the best-fit values found using all
choices of FRR is remarkably good.  On the other hand, DR
significantly underestimates $c_4$.  We can understand the problem
very simply; it is not possible to accurately reproduce the necessary
$1/m_\pi^2$ behaviour of the chiral loops (for $m_\pi > \Lambda$) with
a 3rd order polynomial in $m_\pi^2$.

It is clear that the use of an EFT with a FRR enables one to make an
accurate extrapolation of the nucleon mass as a function of the quark
mass. Indeed, all of the FRR used yield physical nucleon masses that lie 
within a range of 1\%.  
It is especially interesting to observe the very small difference
between the physical nucleon masses obtained with each FRR when we
go from LNA to NLNA -- i.e., when the effect of the $\Delta$ is
included. (The change is typically a few MeV for a FRR but more than 100
MeV for DR.) Once again the convergence properties of the FRR expansion are
remarkable. 
\begin{table}
\begin{center}
\begin{tabular}{lccc}
  & LNA   & \multicolumn{2}{c}{NLNA} \\
Regulator	& $m_N$ & $m_N$ & $\sigma_N$   \\
\hline
Dim.~Reg.	& $0.784$ & $0.884\pm 0.103$  & $50.3\pm 10.0$ \\
Dim.~Reg.~(BP)	& $0.784$ & $0.923\pm 0.103$  & $42.7\pm 10.0$ \\
Sharp cutoff	& $0.968$ & $0.961\pm 0.116$  & $34.0\pm 13.0$ \\
Monopole	& $0.964$ & $0.960\pm 0.116$  & $33.0\pm 13.0$ \\
Dipole		& $0.963$ & $0.959\pm 0.116$  & $33.3\pm 13.0$ \\
Gaussian	& $0.966$ & $0.960\pm 0.116$  & $33.2\pm 13.0$
\end{tabular}
\caption{The nucleon mass, $m_N$ (GeV), and the sigma commutator,
$\sigma_N$ (MeV), extrapolated to the physical pion mass obtained in a
NLNA (fourth order) chiral expansion -- from 
Ref.~\protect\cite{Leinweber:2003dg}.  Convergence of the expansion is
indicated by the nucleon mass obtained in an analysis where we retain
only the LNA (third order) behaviour.
\label{tab:physical}}
\end{center}
\end{table}

\section{Variation of $M_N$ Under a Chiral Invariant Scalar Field}

In order to avoid large violations of chiral symmetry, it is most 
natural to suppose that in models such as QMC we are in a representation 
where $\bar{\psi}_N \psi_N$ is chiral invariant.  For example, if one views 
the scalar, intermediate range NN interaction as the result of two 
pion exchange this is exactly what one would find.  Within such a framework 
the mass of the pion is protected by chiral symmetry~\cite{Bentz:cw},  
even in medium, and one expects only very small deviations of 
the in-medium pion mass from its free value - at least for 
densities typical of normal nuclei.  
This theoretical expectation is certainly supported by recent pionic 
atom data~\cite{Kienle:hq}.

In order to draw conclusions about the scalar polarizability of the 
nucleon the issue to be addressed is then simply: how does the mass 
of the nucleon change when the quark mass varies, 
$m_q \rightarrow m_q^* \equiv m_q - g_\sigma^q \sigma$, 
with the pion mass kept fixed?  
The extensive studies of the nucleon mass in lattice QCD and 
especially the success in relating full and quenched lattice QCD 
simulations~\cite{Young:2002cj}, give us confidence that we understand 
the pion loop contributions. In particular, it was found in Ref.~\cite{Young:2002cj} that 
the coefficients $a_i$ were the same, within fitting errors, in full and quenched QCD 
when we used a dipole regulator with mass 0.8 GeV -- in the case of both the N and 
the $\Delta$. We note that the separation of the pion loop contributions is model 
dependent, unlike the extraction of the renormalized coefficients, $c_i$, where 
we were careful to establish the model independence at the 1\% level. However, the 
possibility of establishing a connection between QCD and fundamental properties of 
nuclear systems, such as saturation, is so important that it seems worthwhile to 
tolerate a degree of model dependence at this stage. 

The free nucleon mass expansion in free space:
\begin{equation}
M_N = a_0 + a_2  m_\pi^2 + a_4 m_\pi^4 + \textrm{self-energy} (m_\pi,\Lambda)
\end{equation}
then becomes
\begin{equation}
M_N^* = a_0 + a_2 { m_\pi^*}^2 + a_4 {m_ \pi^*}^4 + 
\textrm{self-energy} (m_\pi, \Lambda)
\end{equation}
in-medium. As noted above, we use a dipole regulator of 
mass 0.8 GeV for the pion loops, 
which do not change in-medium because the pion mass is protected by chiral symmetry.
To lowest order the GMOR relation gives 
${m_\pi^*}^2 \propto m_q^*$.  However, since the scalar potential is not 
necessarily small, we need to keep the next order term in order 
to extract the dependence 
of the nucleon mass in-medium on the effective quark mass. 

In $\chi$PT the next term is non-analytic in $m_q$~\cite{Gasser:1984gg}.  
However, over the 
relevant mass range, fits to lattice data, as well as fits to 
Schwinger-Dyson phenomenology~\cite{Maris:1999nt,Maris:1997tm} 
and a best fit to the $\chi$PT expansion, 
all suggest that
\begin{equation}
m_\pi^2 \, \, \approx \, \, 4 \, m_q + 20 \, m_q^2,
\end{equation}
with all masses in GeV.

If we now replace the quark mass, $m_q$, by 
$m_q^* \equiv m_q - g_ \sigma^q \sigma$, to second order in $\sigma$ 
one finds $M_N^* = M_N - (4 a_2 g_\sigma^q) \sigma + (20 a_2 + 16 a_4) 
(g_\sigma^q)^2 \sigma^2$, which can be written as $M_N^* -M_N \approx 
g_\sigma (1 - g_\sigma \sigma) \sigma$, if we use the numerical values of 
$a_{2,4}$ from the analysis reported in Sec.~2.2 (without the tadpole). 
That is, $g_{\sigma \sigma} \approx - g_\sigma^2$ and 
the reduction of the effective 
$\sigma$-nucleon coupling at nuclear matter density, where $g_\sigma \sigma 
\in (150,200)$ MeV is around 15-20 \% .  This is in remarkable agreement with 
the upper end of the values found phenomenologically in the QMC model.

If instead of the values found in Sec. 2.2 one used the more recent fit 
of Table~\ref{tab:afit} (including the tadpole term), the coefficient of 
$g_\sigma^2$ in the expression for $g_{\sigma \sigma}$ 
would be -0.6 instead of -1. This is in remarkable agreement with the value 
given by the bag model ($g_{\sigma \sigma} = -0.11 R g_\sigma^2$, with $R=1 {\rm fm} =
5 {\rm GeV}^{-1}$). 
In terms of the phenomenological role of $g_{\sigma \sigma}$ in nuclear saturation, 
the difference between these two estimates is unimportant.

To summarize, the FRR expansion of the nucleon mass in terms of quark mass, 
with parameters determined by fits to the best two-flavor, lattice 
QCD simulations, leads to a scalar polarizability of the sign and 
magnitude found in the QMC model.  Amongst the natural consequences of 
this result, as we explained in the Introduction, is that nuclear matter will
naturally saturate as the density increases. Thus in a very meaningful sense 
the present result connects the saturation of nuclear matter to the 
structure of the nucleon itself within QCD.

\section{Conclusion}

The remarkable progress in resolving the problem of chiral extrapolation of lattice QCD 
data, in the case (such as the nucleon mass) where the chiral coefficients are known 
accurately, gives one confidence that the pion loop contributions are under control. 
In the case of the nucleon one can use this control to estimate the effect of applying 
a chiral invariant scalar field to the nucleon -- i.e. to estimate the so-called 
``scalar polarizability'' of the nucleon. We stress that, unlike the determination of 
low energy coefficients, which is demonstrably model independent, for the present the 
determination of the scalar polarizability requires a choice of regulator. We chose 
a dipole of mass 0.8 GeV because of the connection between full and quenched QCD 
that has been established in this case. The resulting value is in excellent 
agreement with the range found in the quark-meson coupling model (QMC), which has 
provided an impressive description of many phenomena in nuclear physics. Of most 
immediate significance is the fact that such a scalar polarizability very naturally 
yields nuclear saturation within a relativistic mean field treatment of nuclear matter.
Thus in a very real sense the results presented here provide a direct connection 
between our growing power to compute hadron properties from QCD itself and fundamental 
properties of atomic nuclei. This is especially interesting in the light of recent 
experimental evidence~\cite{Strauch:2002wu,Dieterich:2000mu,Fissum:2004we} 
that the structure of a bound ``nucleon'' differs from that 
observed when it is free~\cite{Lu:1998tn,Lu:1997sd,Thomas:1998eu}. 
These are important problems that will be more and more 
a focus for our field in the coming years.

\section*{Acknowledgements}
This work is supported by the Australian Research Council and by DOE  
contract DE-AC05-84ER40150, under which SURA operates 
Jefferson Laboratory.

%

\end{document}